\documentclass[%
aps,
 sd,%
 superscriptaddress,
 amsmath,amssymb,
twocolumn,%
]{revtex4-1}

\usepackage{graphicx}
\usepackage{dcolumn}
\usepackage{bm}
\usepackage{upgreek}

\begin{document}

\preprint{APS/123-QED}

\title[Injection Locking of Quantum-Dot Microlasers Operating in the Few-Photon Regime]{Injection Locking of Quantum-Dot Microlasers Operating in the Few-Photon Regime}

\author{Elisabeth Schlottmann}
\thanks{These authors contributed equally to this work.}
\affiliation{Institut f\"ur Festk\"orperphysik, Quantum Devices Group, Technische Universit\"at Berlin, \\
Hardenbergstra{\ss}e 36, EW 5-3, 10623 Berlin, Germany}

\author{Steffen Holzinger}
\thanks{These authors contributed equally to this work.}
\affiliation{Institut f\"ur Festk\"orperphysik, Quantum Devices Group, Technische Universit\"at Berlin, \\
Hardenbergstra{\ss}e 36, EW 5-3, 10623 Berlin, Germany}

\author{Benjamin Lingnau}
\affiliation{Institut f\"ur Theoretische Physik, AG Nichtlineare Laserdynamik, Technische Universit\"at Berlin,
Hardenbergstra{\ss}e 36, EW 7-1, 10623 Berlin, Germany}

\author{Kathy L\"udge}
\affiliation{Institut f\"ur Theoretische Physik, AG Nichtlineare Laserdynamik, Technische Universit\"at Berlin,
Hardenbergstra{\ss}e 36, EW 7-1, 10623 Berlin, Germany}

\author{Christian Schneider }
\affiliation{Technische Physik, Universit\"at W\"urzburg, Am Hubland, 97074 W\"urzburg, Germany}

\author{Martin Kamp}
\affiliation{Technische Physik,  Universit\"at W\"urzburg, Am Hubland, 97074 W\"urzburg, Germany}

\author{Sven H\"ofling}
\affiliation{Technische Physik, Universit\"at W\"urzburg, Am Hubland, 97074 W\"urzburg, Germany}
\affiliation{SUPA, School of Physics and Astronomy, University of St Andrews, St Andrews, KY16 9SS, United Kingdom}

\author{Janik Wolters}
\thanks{janik.wolters@tu-berlin.de}
\affiliation{Institut f\"ur Festk\"orperphysik, Quantum Devices Group, Technische Universit\"at Berlin, \\
Hardenbergstra{\ss}e 36, EW 5-3, 10623 Berlin, Germany}

\author{Stephan Reitzenstein }
\affiliation{Institut f\"ur Festk\"orperphysik, Quantum Devices Group, Technische Universit\"at Berlin, \\
Hardenbergstra{\ss}e 36, EW 5-3, 10623 Berlin, Germany}

\date{\today}

\begin{abstract}

We experimentally and theoretically investigate injection locking of quantum dot (QD) microlasers in the regime of cavity quantum electrodynamics (CQED). We observe frequency locking and phase-locking where cavity enhanced spontaneous emission enables simultaneous stable oscillation at the master frequency and at the solitary frequency of the slave microlaser. Measurements of the second-order autocorrelation function prove this simultaneous presence of both master and slave-like emission, where the former has coherent character with $g^{(2)}(0)=1$ while the latter one has thermal character with $g^{(2)}(0)=2$.
Semi-classical rate-equations explain this peculiar behavior by cavity enhanced spontaneous emission and a low number of photons in the laser mode.


\end{abstract}





\maketitle
\section{INTRODUCTION}
Controlling an oscillator's frequency by injecting external signals is a universal concept in nonlinear science and applies to a multitude of physical and biological systems~\cite{Pikovsky2008,Anishchenko2000,Rippard2005, Wilke2002}.
It is an integral part of mobile communication or digital video broadcasting. Beyond various applications 
of injection locking there has been enormous interest in the physical understanding of this important phenomenon, and for conventional macroscopic oscillators it is well understood by Adler's theory~\cite{Adler1946,Erneux2010}. In contrast, injection locking in microsystems operating close to the quantum level of light and matter is an almost unexplored field of research in which only few experiments have been realized~\cite{Astafiev2007, Andre2009}. Indeed, the question appears how miniaturized devices, such as micro- or nanolasers operating in the few photon regime, respond to the external stimulus by the master laser. \\
Micro- and nanolaser have been studied extensively in recent years and are usually based on dielectric or plasmonic laser resonators~\cite{Oulton2009, Lu2012, Painter1999, Khaja2012, Hill2007} with an effective 
mode volume on the order of the cubic wavelength. As a result light-matter coupling is strongly enhanced~\cite{Vahala2003} and spontaneous emission coupling factors ($\beta$-factors) close to unity are observed \cite{Lermer2013}. This allows for lasing with a greatly reduced threshold and only a few tens of photons in the laser mode~\cite{Strauf2006, Reitzenstein2006, Nomura2010}. In recent experiments such lasers operating in the regime of cavity quantum electrodynamics (CQED) have been used to investigate nonlinear effects such as spontaneous chaos induced by delayed feedback~\cite{Albert2011} or symmetry breaking~\cite{Hamel2015}. \\
In this letter, we study the external control of a CQED enhanced microlaser exhibiting an ultra-low threshold and less than 100 photons in the cavity. Where synchronization of the slave laser to the master frequency is expected for its conventional counterparts, stationary oscillation synchronized to the external signal and oscillation at the solitary frequency is observed to occur simultaneously for our microlaser. As evidenced by a theoretical analysis, the observed partial injection locking is a phenomenon unique to nonlinear single mode high-$\beta$ microlasers excited with a few tens of quanta. 
Interestingly, the remaining emission at the solitary frequency of the slave microlaser shows thermal character associated with an autocorrelation function $g^{(2)}(0) = 2$. 
\section{ DEVICES UNDER STUDY}
The CQED enhanced lasers used in our experiments are micropillar lasers, with a single layer of In$_{0.3}$Ga$_{0.7}$As quantum dots with an area density of $5\cdot 10^9 /$cm$^2$ acting as a gain medium, sandwiched between AlAs/GaAs distributed Bragg reflectors forming a $\lambda$-cavity. Using electron beam lithography and plasma etching pillar structures with a diameter of 5.3\ $\mu$m are fabricated, planarized with benzocyclobutene, and contacted with ring shaped gold contacts. From the linewidth at inversion the quality factor is estimated to $Q =21\,000$. 
Cf. Ref.~\cite{Boeckler2008} for details on the sample fabrication.
Due to unavoidable structural asymmetries the degeneracy of the two orthogonal polarized fundamental cavity modes is lifted~\cite{Reitzenstein2007}. In the present device this leads to a complete suppression of the higher energy mode and lasing was observed on the lower energy mode only. The device operates as a single mode laser.
\\
\begin{figure*}[htb]
\center
\includegraphics[width=1\textwidth]{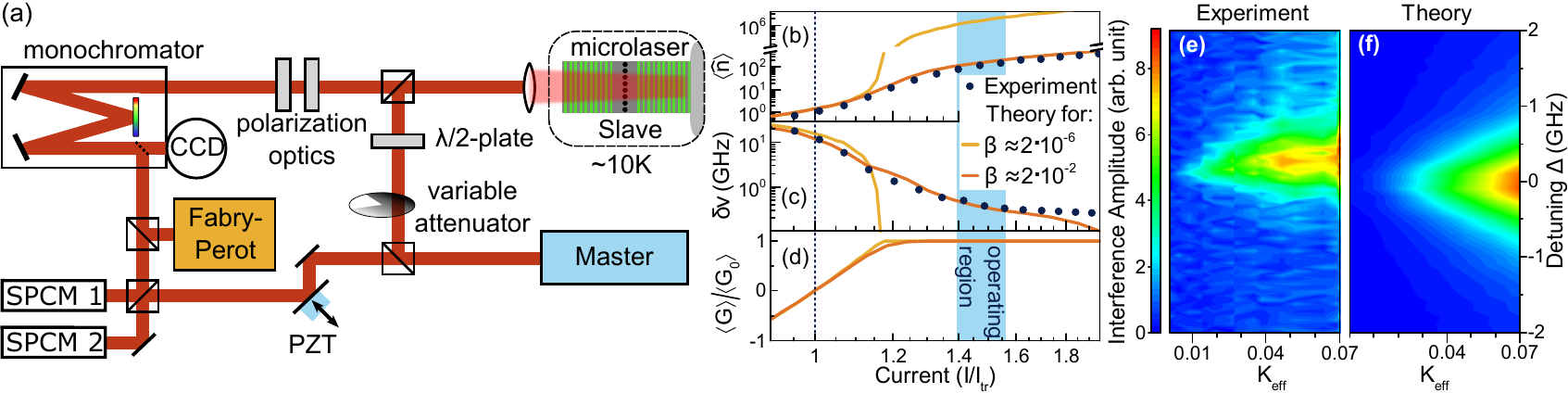}
\caption{\textbf{ The microlaser and its optical characteristics} (a) Sketch of the optical setup used in the experiments. (b)-(c) Average photon number $\langle \hat n \rangle$ and linewidth $\delta\nu_{0}$ (FWHM) obtained from the measured output characteristics and numerical simulations respectively as function of normalized current $I/I_{thr}$. The investigated CQED microlaser exhibits $\beta\sim 10^{-2}$, while for comparison the calculation for $\beta\sim10^{-6}$ corresponds to a conventional low-$\beta$ laser. (d) Calculated normalized gain $\langle  G \rangle=2\, Z \, g_{0}^{2}\left(2\langle \hat \rho \rangle -1\right)$. (e) Amplitude of the phase-locked slave oscillation as a function of detuning $\Delta$ between master and slave and injection strength $K_{eff}$, measured at high electrical current. (f) Simulation result corresponding to (e).
}
\label{fig:setup}
\end{figure*}
The sample is mounted in a continuous flow He-cryostat at a temperature of $T \sim$~15~K, and electrically pumped by a precision current source. Emission is collected with an aspherical lens (NA=0.5), spatially filtered and analyzed with a power-meter, a spectrometer (spectral resolution 6.5~GHz) or alternatively with a scanning Fabry-P\'{e}rot cavity ($7.5$~GHz free spectral range, 100 MHz resolution). 
The cavity photon numbers $\langle\hat n_{0} \rangle$ are deduced from the numerical simulations of the CQED microlaser and are in agreement with the measured emission power of $P_{S}=h\nu_{0} \eta \kappa \langle \hat n_{0} \rangle = 470\,$nW at $I=1.4 \,I_{thr}$ with detection efficiency $\eta\sim0.85$ for the directed laser emission. \\
\section{THEORETICAL MODEL}
The microlaser is modeled by semi-classical rate equations based on a quantum Langevin approach to the CQED system in the weak coupling regime. 
This allows for a simple, but accurate incorporation of the phase-sensitive injection of the master field $E_{M}=|E_{M}|e^{i 2\pi\Delta t}$. 
The resulting differential equations for the complex electric field in the cavity $E$, where the photon number is given by  $\langle \hat n\rangle=|E|^{2}/\mathcal{E}_{0}^{2 }$, the average quantum dot occupation $ \rho=\langle \hat \rho(t)\rangle$ and the wetting layer occupation  $w=\langle \hat w(t)\rangle$ read:
\begin{eqnarray}
\dot E  &= &\left[2Z g_{0}^{2}\left(2 \rho -1\right) - \frac{\kappa}{2}  \right] (1+i\alpha) E  + \frac{\kappa}{2} E_\mathrm{M} +F_{S}, \\
\dot \rho &=& -  \frac{ g_{0}^{2}}{\mathcal{E}_{0}^{2}}|E|^2 \big(2\ \rho  -1\big) + S w \big(1-\rho \big) - \frac{ \rho  }{\tau_{QD}} ,\label{eq:qd}\\
 \dot w &=& \frac{I}{e_0 A} - \frac{2Z}{A} S w \big(1-\rho\big) - \frac{2Z^\mathrm{inact}}{A} \frac{\rho^\mathrm{inact}}{\tau_{QD}} - \frac{w}{\tau_w},
\end{eqnarray}
where from the total number of QDs in the cavity a portion $Z=250$ is considered as active and $Z^\mathrm{inact}=750$ is considered as inactive \cite{Redlich2016}, but pumped by the injected current. $\mathcal{E}_{0}=\sqrt{\frac{h \nu_{0}}{2\varepsilon_{r} \varepsilon_0 V}}$ is the electric field per cavity photon, with $\varepsilon_0$ and $\epsilon_r=11$ denoting the vacuum permittivity and background dielectric constant, respectively. $A=22\upmu$m$^2$ and $V=4\upmu$m$^3$ are the 
area of the pillarÕs cross-section and the mode volume, respectively. The photon loss rate is given by $\kappa=\frac{2\pi \nu_{0}}{Q}=100$ns$^{-1}$. The amplitude-phase coupling coefficient $\alpha=2$ was extracted from the experimental above-threshold linewidth shown in Fig. \ref{fig:setup} \cite{Wieczorek2005}, and $E_\mathrm{M}$ is the external master laser field incident at the micropillar. 
 The Langevin source term $F_{S} = \mathcal{E}_{0} \sqrt{\frac{2\beta Z \langle \hat \rho\rangle }{\tau_{QD}}} \xi(t)$ models spontaneous emission, with 
a Gaussian white noise  $\xi(t)$. 
The Purcell-reduced spontaneous lifetime in the QDs  $\tau_{QD}=110$~ps and in the wetting layer  $\tau_w=1$~ns, respectively, which are determined from fits to the input-output curve, Fig. \ref{fig:setup} (b-c). All values are in agreement with measured values in similar devices \cite{Reitzenstein2012}.  The occupation of the inactive dots is calculated from the static solution of Eq.~\eqref{eq:qd} without stimulated recombination, which yields $\rho^\mathrm{inact} = \frac{\tau_{QD} S w}{1 + \tau_{QD} S w }$.

For our experimental studies a quantum dot microlaser 
with emission frequency $\nu_{0}= 352$~THz was used. 
The  measured input/output characteristics presented in Fig. \ref{fig:setup} (b, c) were reproduced quantitatively by our numerical model. According to the simulations, the onset of stimulated emission occurs at a current of $I_{thr} = 24\; \upmu$A. The $\beta$-factor is extracted from the fits to the input-output curves of the micropillar laser along with its sub-threshold linewidth, which both depend sensitively on the choice of beta. We extract $\beta \sim 2\%$.
The investigated laser shows a rather smooth transition from spontaneous to stimulated emission. This behavior is typical for CQED enhanced microlasers~\cite{Gies2007}, and stands in contrast with conventional lasers with  $\beta \gtrsim10^{-5}$ exhibiting a drastic increase of emission intensity above threshold. Similar, linewidth reduction and gain clamping are less pronounced compared to their conventional counterparts (Fig. \ref{fig:setup} b-d). \\
\begin{figure*}[hbt]
\center
\includegraphics[width=0.99\textwidth]{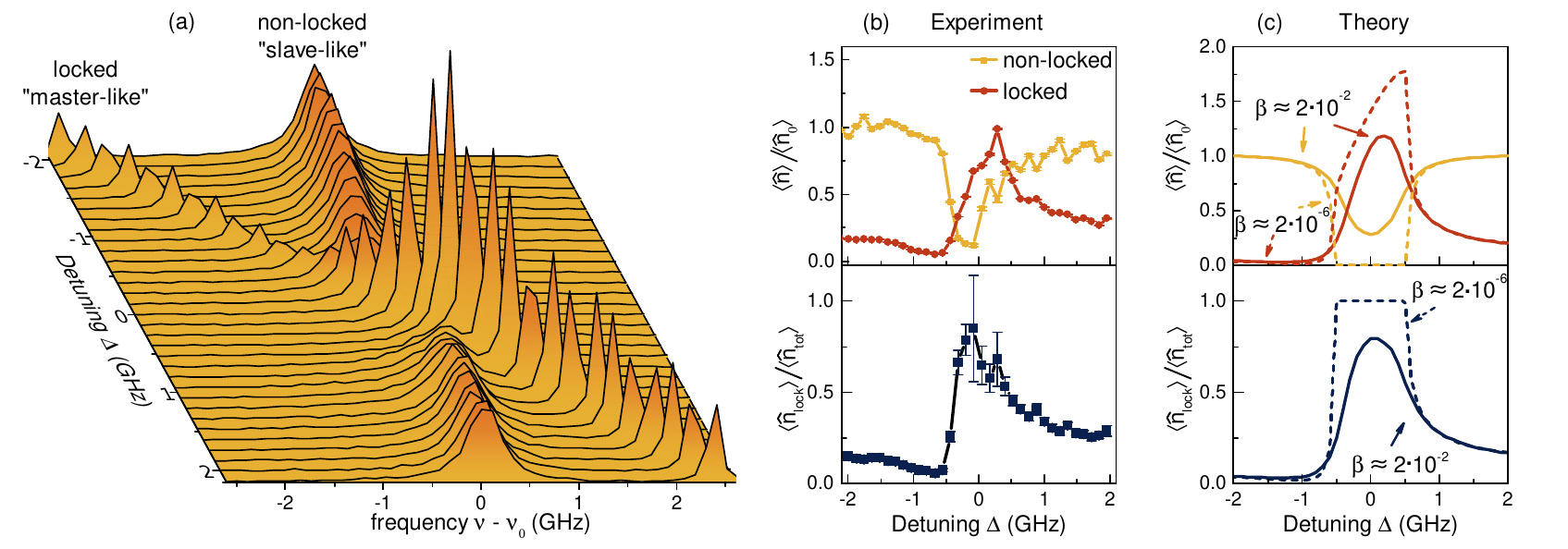}
\caption{\textbf{Partial injection locking of the QD-microlaser} (a) High resolution spectra of the emission at $I=1.6\, I_{thr}$ under optical injection with $K_{eff}$ $\sim$ 0.17 for various detunings.  The frequency axis is relative to the free running slave frequency of $\nu_{0}= 352$~THz. The microlaser shows a large region of partial injection locking, where phase-locked oscillation at the master frequency  $\nu_{0}+\Delta$ and non-locked oscillation at the slave frequency $\nu_{0}$ occur simultaneously. (b) Upper panel: Intensity of the phase-locked (master-like) and the non-locked (slave-like) oscillation as obtained from the measurements shown in (a). All error bars correspond to one standard deviation as obtained from fitting the data with Lorentzian curves. Lower panel: Ratio between locked and total oscillation intensity, indicating that despite the comparable large injection strength, perfect locking is not achieved.  (c) Numerical simulations corresponding to (b). For the used microlaser with $\beta\sim2\%$ (solid lines) partial locking is reproduced, while for a conventional laser with $\beta = 0$ (dashed lines) complete locking is predicted. }
\label{fig:LockingMap_FPI}
\end{figure*}
\section{INJECTION EXPERIMENTS}
For a first injection experiment, a rather high current of $I=2.3 \,I_{thr}$ was chosen. 
As master oscillator,  emission from a tunable external cavity diode laser with frequency $\nu_{M}$ was injected into the CQED oscillator. An upper limit for the injection strength $K_{meas} = \sqrt{P_M/P_S}$ was obtained by measuring the free-running slave emission power $P_S$ and the reflected master power $P_M$ in absence of current $I$ after spatial filtering by a nearly diffraction limited pinhole. 
Due to imperfect mode matching, the injection efficiency is below unity. We extract the effective injection ratio $K_{eff}$ from numerical fits to the injection locking dynamics (Fig. 1(f) and Fig. 4(b), respectively). Best agreement between experiment and theory is obtained for $K_{eff} = K_{meas}/10$.Ó
To prove phase-synchronization between master laser and CQED laser, emission from both was interfered on a fiber beam-splitter. In this configuration, the amplitude of the interference fringes obtained by varying the master's phase with a dithering mirror corresponds to the phase-locked oscillation amplitude of the slave. 
Fig. 1 (e) shows a map of this amplitude, depending on injection rate $K$ and detuning between master and slave \mbox{$\Delta=\nu_{M}-\nu_{0}$}. A slightly asymmetric locking cone indicates phase-locking up to $\Delta\sim$~1.5~GHz. 
The asymmetry of the locking cone originates in amplitude-phase coupling being common for semiconductor lasers~\cite{Henry1982, Lingnau2014} and interference with directly reflected master radiation for larger values of $K$. Towards the border of the locking region, the slave oscillation amplitude decreases, as the oscillator is forced to oscillate at an unpreferable frequency. In this regard, the microlaser behaves under external injection similar to conventional lasers, where 
 the locking range is given approximately by $\frac{K_{eff} \nu_{0}}{Q}\sqrt{1+\alpha^2 } \sim 1~$GHz for $K_{eff}=0.03$~\cite{Wieczorek2005,Erneux2010}. \\
For experiments in the ultra-low light level regime, the region of stable single mode laser oscillation with $I\sim1.5~I_{thr}$ was chosen, where the measured linewidth  is $\delta\nu_{0}\sim350$~MHz. In this regime the average photon number in the cavity is $\langle \hat n_{0} \rangle\sim 100$, at least three orders of magnitude lower compared to conventional lasers. In addition to phase-sensitive measurements, high resolution spectra were recorded for varying  $\Delta$ at  $K_{eff}= 0.17$ (Fig.~\ref{fig:LockingMap_FPI}). For positive  $\Delta$, strong phase-locked emission at the master frequency is visible, while weaker phase-locked emission is measured for negative detunings $\Delta$ as expected from the interference measurements. 
Surprisingly, this is accompanied by strong non-phase-locked emission with the slave's solitary frequency and linewidth, indicating a regime of \textit{partial locking} where the laser oscillates at the master frequency \textit{and} at its solitary frequency simultaneously. 
Only for detunings smaller than the emission linewidth of the solitary laser, $|\Delta| < \delta\nu_{0}$, the spectrally broad non-locked emission is suppressed and narrow-band phase-locked emission becomes dominant. 
This phenomenon of partial injection locking is well reproduced by numerical simulations taking into account the impact of Purcell-enhanced spontaneous emission in the high-$\beta$ laser operating with few photons in the cavity.
 Interestingly, partial injection locking is not theoretically predicted for conventional low-$\beta$ single mode lasers. We would also like to note that in lasers with a more complex mode structure, an intensity reduction (but not complete suppression) under optical injection has been observed. 
When addressing injection in higher-order modes of VCSELs~\cite{Hong2002} or in a different polarization mode~\cite{Valle2007, Gatare2006}, a suppression of one laser mode emission was achieved by gain competition with other modes of the laser. In contrast, our experiments are performed with a single mode laser. Here, in the locking regime the emission at the solitary slave laser frequency is mainly driven by CQED enhanced spontaneous emission. \\
\vspace{-0.5cm}
\begin{figure}[htb]
\center
\includegraphics[width=1\columnwidth]{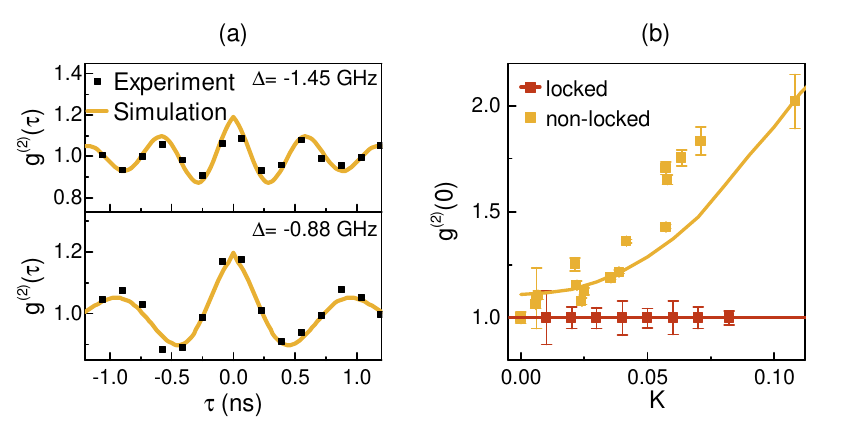}
\caption{\textbf{Signatures of partial injection locking in the time domain} (a) Time-dependent second-order autocorrelation measurements for $\Delta=-1.45$~GHz (top) and  $\Delta=-0.88$~GHz (bottom). Both show an exponentially decaying oscillatory behavior, caused by the beat note between the simultaneously present spectrally narrow locked and the spectrally broad non-locked oscillation modes. (b) Frequency-selectively measured photon statistics of the locked and non-locked emission, with absolute value of $\Delta$ comparable to the lower panel in (a). The latter one undergoes a transition from coherent to thermal emission $g^{(2)}(0)=2$ with increasing injection strength $K_{eff}$. Solid lines are numerical simulations.}
\label{fig:g2_cohTime_Full}
\end{figure} \\
\begin{figure*}[ht]
\center
 \includegraphics[width=0.99\textwidth]{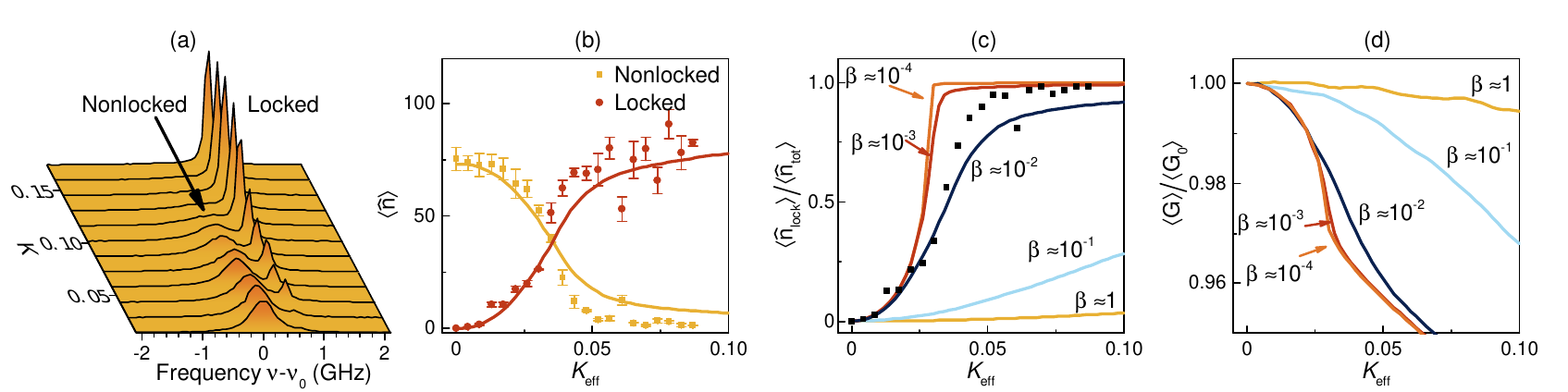}
\caption{\textbf{Mode intensities and gain under variation of the injection strength with $\Delta=500\,$MHz} (a) High-resolution emission spectra of the CQED microlaser at $I=1.4\, I_{thr}$ under optical injection. Locked oscillation at frequency $\nu_{0}+\Delta$ and non-locked oscillation near frequency $\nu_{0}$ is visible within a large range of $K_{eff}$ values.
(b) Intensity of locked (red) and non-locked (yellow) oscillation for varying $K_{eff}$. Dots are measured values extracted from (a), while solid lines correspond to numerical simulations of the device.  (c) Contribution of the locked emission to the overall emission $\langle\hat n_{lock}\rangle / \langle \hat n_{tot} \rangle$ for various $\beta\sim 1 -10^{-4}$ corresponding to $\langle \hat n_{0}\rangle \sim 5-7000$. Dark blue squares correspond to the measurement shown in (b). (d) Calculated gain $\langle G\rangle$ as function of $K_{eff}$, corresponding to the $\beta$-values in (c).}
\label{fig:secondarypeakintensity}
\end{figure*}
To gain insight into the dynamical processes, measurements of the second-order photon autocorrelation function were performed with a Hanbury-Brown and Twiss (HBT) setup \cite{Hanbury1956}. 
As expected for the master oscillator and equally for the slave without injection the second order autocorrelation function is constant $g^{(2)}(\tau)=1$. 
In contrast, under optical injection an oscillatory correlation function is observed (Fig.~\ref{fig:g2_cohTime_Full}a).
This feature is explained by partial injection locking, where the simultaneous emission at two frequencies leads to a beat note observed in the time domain.
The frequency of the measured oscillation corresponds exactly to the frequency difference between phase-locked and non-locked oscillation, while the exponential decay of its amplitude is related to the spectral linewidth of the two oscillation modes.
In agreement with the simulations, these measurements prove that the presence of a phase-locked and non-locked components in the spectra of Fig.~2a is not caused by (fast) chaotic switching \cite{Redlich2016} as it can occur in deterministic nonlinear oscillators under external injection \cite{Goulding2007, Lingnau2012}, but the CQED microlaser indeed oscillates in a superposition of both oscillation modes. \\
The photon statistics can be determined by performing HBT measurements spectrally filtered by a Fabry-P\'{e}rot cavity for both frequencies separately. The locked emission shows as expected coherent emission ($g^{(2)}(0)=1$) independent of the injection strength $K_{eff}$ (Fig.~\ref{fig:g2_cohTime_Full}(b)). However, the non-locked emission exhibits an increasing $g^{(2)}(0)$ with increasing $K_{eff}$ and finally reaching the thermal limit $g^{(2)}(0)=2$. 
Thus, optical injection in a microlaser provides the unique and appealing opportunity of controlling the photon statistics at the slaves solitary frequency by gradually shifting the contribution of stimulated emission to the master lasers frequency. \\
For further insight into the physical origin of partial injection locking, we adjust the injection current to $I=1.4\, I_{thr}$, where the laser operates with only 70 photons in the cavity. While keeping the detuning constant at  $\Delta = \delta\nu_{0}$, the injection strength $K_{eff}$ was varied. For conventional nonlinear oscillators without quantum noise an abrupt onset of synchronization above a critical injection strength $K_{crit}$ is expected. In addition, close to $K_{crit}$ the external injection may lead to chaotic dynamics via gain competition effects~\cite{Weiss1991}. For the given parameters of our microlaser $K_{crit}\sim 0.1$ is expected~\cite{Lingnau2012}. In contrast to the expectations for conventional lasers, we find a rather smooth transition from free running to synchronized oscillation when experimentally increasing $K_{eff}$  (Fig.~4). In the transition region phase-locked oscillation at the master frequency $\nu_{0}+ \Delta$ and non-locked oscillation near  $\nu_{0}$ occur simultaneously, comparable to the case of varying detuning. Again, this effect is not attributed to chaotic switching between different oscillation modes observed in conventional systems: second-order autocorrelation measurements indicate constant stationary oscillation. Chaotic dynamics are suppressed in our experiments by quantum fluctuations~\cite{Sattar2016, Wimberger2014, Weiss1991}, as recently predicted for optomechanical systems~\cite{Bakemeier2015}. \\
The used quantum mechanical rate equation model perfectly describes the phenomenon of partial injection locking and the resulting dynamics.
Figure~\ref{fig:secondarypeakintensity}(c) shows the ratio between the average photon number in the phase-locked mode and both oscillation modes for various values of $\beta$ between unity and zero. For conventional macroscopic laser oscillators exhibiting small $\beta$-factors and large average photon numbers, injection with $K_{eff} > K_{crit}$ leads to an efficient suppression of non-locked oscillations and thus a ratio of 1. For the investigated CQED oscillator with $\beta\sim10^{-2}$ the non-locked oscillation intensity is comparable to the locked intensity for a wide range of parameters.
With increasing $\beta$, the presence of partial injection locking gradually increases, while in the limit of $\beta \sim 1$, i.e. thresholdless lasers, the non-locked emission is dominant even for large injection strength. In such devices injection locking is impeded by spontaneous emission noise.
This partial injection locking in CQED microlasers is a genuine phenomenon attributed to Purcell enhanced spontaneous emission noise playing a dominant role in a system with only a few tens of photons in the cavity. \\
The rate equation model gives insight into the underlying physical mechanism:
Injection of the master oscillator effectively reduces the cavity loss rate $\kappa$ for phase-locked oscillation. Thereby, the threshold for oscillations at the master frequency $I_{thr} \sim \kappa$ is reduced and consequentely the gain $\langle  G \rangle$ is clamped to a lower value compared to the free running oscillator (Fig.~4d). 
In conventional lasers with low $\beta$-factor and a number of cavity photons well above $10^3$ this mechanism suppresses non-locked oscillation, even for very small $K_{eff}$. 
In contrast, in CQED microlasers the input/output characteristics are rather smooth and gain competition is less pronounced. \\
\section{CONCLUSIONS}
In conclusion, we have studied  optical injection on a quantum dot microlaser operating in the regime of cavity quantum electrodynamics with $70-150$ photons in the cavity. We found the surprising phenomenon of \textit{partial injection locking}, where complete synchronization would be expected from conventional lasers with three orders of magnitude higher photon numbers. In this regime of partial injection locking, the laser oscillates phase-locked to the master and non-locked simultaneously. This effect is not predicted by classical deterministic theories of single-mode nonlinear oscillators, but is understood by the quantum theory of lasers including CQED enhanced spontaneous emission. Measurements of the photon statistics reveal thermal emission associated with $g^{(2)}(0)=2$ of the non-locked oscillations at high injection strength. 
Our results pave the way for further studies on the dynamics of CQED enhanced microlasers and explores the limits of external quantum control of nanophotonic systems.  As such, it connects to possible applications 
e.g. in optical data communication, where injection-locking can increase the modulation bandwidth of conventional lasers \cite{Simpson1997}. Furthermore, chaotic dynamics induced by feedback injection is applied in secure chaos communication and random number generation \cite{Sciamanna2015}.
When using miniaturized and integrated devices, our work is highly relevant for both applications.
Most importantly, our work will enable reservoir reservoir computing with optically coupled laser arrays \cite{Larger2012,Brunner2013}.
This type of neuromorphic computing is based on diffractive coupling of lasers and crucially relies on optical injection as coupling mechanism. 
Strong benefit is expected from dense arrays of microlasers in order to realize large networks of coupled lasers. Hence, exploration and understanding of optical injection 
in microlasers directly connects to important applications in the fields of advanced optical data communication and optical processing.

\begin{acknowledgments}
The research leading to these results has received funding from the European Research Council under the European
Union's Seventh Framework ERC Grant Agreement No. 615613 and from the German Research Foundation via CRC 787.
\end{acknowledgments}

\end{document}